\documentclass[12pt]{article}

\usepackage[english]{babel}

\usepackage[letterpaper,top=2cm,bottom=2cm,left=3cm,right=3cm,marginparwidth=1.75cm]{geometry}

\usepackage{amsthm,amsmath,amsfonts,amssymb}
\usepackage{graphicx}
\usepackage[colorlinks=true, allcolors=blue]{hyperref}
\usepackage[round]{natbib}
\usepackage{caption}
\usepackage{subcaption}
\usepackage{bbm}
\usepackage{hyperref}
\usepackage{xcolor}
\usepackage{enumitem}
\usepackage{authblk}
\usepackage{setspace}
\newtheorem{proposition}{Proposition}

\usepackage{algorithm}
\usepackage{algorithmic}

\makeatletter
\newcommand{\ALOOP}[1]{\ALC@it\algorithmicloop\ #1%
  \begin{ALC@loop}}
\newcommand{\ENDALOOP}{\end{ALC@loop}\ALC@it\algorithmicendloop}

\makeatother

\title{\bf A new block covariance regression model and inferential framework for massively large neuroimaging data}
\author[1]{Hyoshin Kim}
\author[1]{Sujit K. Ghosh}
\author[1]{Emily C. Hector}
\affil[1]{Department of Statistics, North Carolina State University}
\date{}

\begin{document}
\maketitle

\begin{abstract}
\noindent Some evidence suggests that people with autism spectrum disorder exhibit patterns of brain functional dysconnectivity relative to their typically developing peers, but specific findings have yet to be replicated. To facilitate this replication goal with data from the Autism Brain Imaging Data Exchange (ABIDE), we propose a flexible and interpretable model for participant-specific voxel-level brain functional connectivity. Our approach efficiently handles massive participant-specific whole brain voxel-level connectivity data that exceed one trillion data points. The key component of the model is to leverage the block structure induced by defined regions of interest to introduce parsimony in the high-dimensional connectivity matrix through a block covariance structure. Associations between brain functional connectivity and participant characteristics -- including eye status during the resting scan, sex, age, and their interactions -- are estimated within a Bayesian framework. A spike-and-slab prior facilitates hypothesis testing to identify voxels associated with autism diagnosis. Simulation studies are conducted to evaluate the empirical performance of the proposed model and estimation framework. In ABIDE, the method replicates key findings from the literature and suggests new associations for investigation.
\end{abstract}

\section{Introduction}

Although Autism Spectrum Disorder (ASD) is frequently described as a disorder of dysconnection, there remains no consensus on which specific brain regions exhibit hyper- or hypo-connectivity. Additionally, ASD is highly heterogeneous, meaning that participants with the disorder vary across several dimensions in their clinical presentation, including behavioral deficits, intellectual functioning, sex, executive functioning, and developmental histories \citep{lord2018autism}. Much of the current research seeks to further understand the functional relationships between different brain areas and their connection to clinical phenotypes in ASD.

Resting-state functional MRI (rfMRI) data from the Autism Brain Imaging Data Exchange (ABIDE) repository \citep{DiMartinoetal2014} represent a unique opportunity to replicate existing findings and build consensus among existing results due to their size and scope. The data we consider are preprocessed using the Configurable Pipeline for the Analysis of Connectomes (CPAC) \citep{craddock2013neuro} with a preprocessing strategy that includes band-pass filtering and global signal regression, and registered to a common template. We select the hierarchical multiresolution 17-network parcellation \citep{Schaeferetal2018}, which consists of 200 regions of interest (ROIs) and is widely used in functional connectivity research \citep{hansen2022mapping, wang2024characterization}, including studies of ASD. This parcellation is well-suited for capturing large-scale functional networks while maintaining a suitable spatial resolution of ROIs. Based on this scheme, for each participant $i \in \{1, \dots, n\}$, the $M_{i} \approx 42{,}750$ voxels are grouped into $J = 200$ ROIs, which are themselves organized into 17 brain networks. We consider lag-2 thinned rfMRI time points of length $T_i$ in each voxel, where the thinning process yields observations that are approximately independent for each participant. The participant-specific functional connectivity matrix is computed by calculating the sample correlation between the $M_{i}$ voxels, treating the $T_i$ time points as replicates. In whole-brain analyses, this $M_{i} \times M_{i}$ connectivity matrix, which corresponds to the between-voxel sample correlation matrix, is used to explore patterns of co-activation and reciprocal activation between voxels. The resulting voxel-level dataset consists of $\sum_{i=1}^{n} M_{i}^{2}$ between-voxel correlations, exceeding one trillion entries. This extensive dataset facilitates an unparalleled, high-resolution exploration of brain connectivity, but frequently only summaries are used due to the prohibitive computational and memory needs of analyzing such massive and complex data. 

\begin{figure}[ht!]
    \centering
    \includegraphics[width=0.9\linewidth]{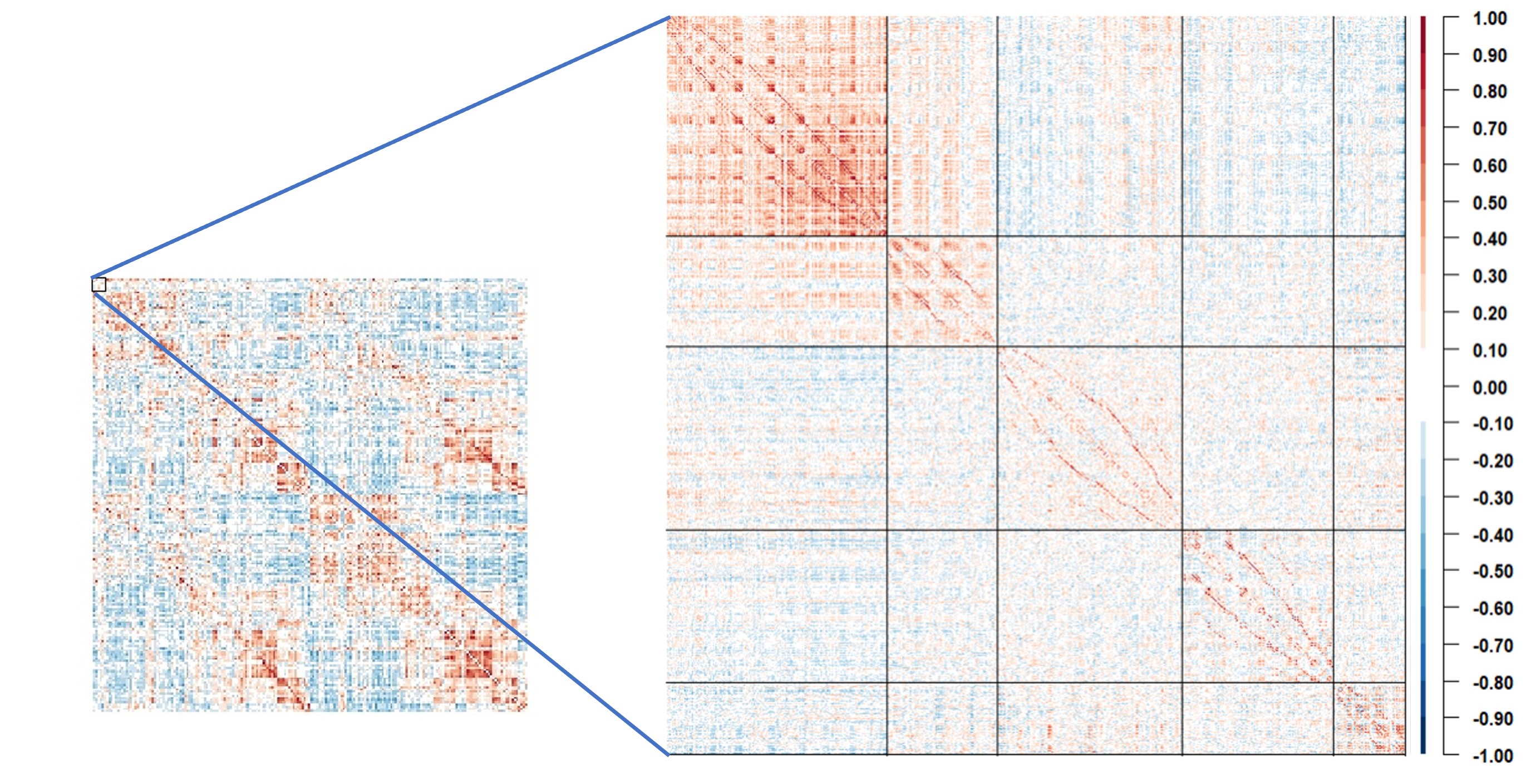}
    \caption{ROI-level functional connectivity matrix (left) and voxel-level functional connectivity matrix for five ROIs (right) for one participant. The first five ROIs are shown based on the \citet{Schaeferetal2018} hierarchical multi-resolution parcellation with $17$ networks and $200$ ROIs. The matrix dimensions are $867 \times 867$, corresponding to the following ROI sizes: $259$, $129$, $216$, $178$, and $85$ voxels. Black horizontal and vertical lines indicate boundaries between ROIs. 
    }
    \label{fig:motiv}
\end{figure}

Figure \ref{fig:motiv} displays the voxel-level connectivity matrix for the first five ROIs of a randomly selected participant in ABIDE, highlighting several key characteristics and challenges associated with handling these data. First, Figure \ref{fig:motiv} shows that connectivity values consist of a mix of positive and negative correlations, which complicates the modeling process. Flexibly capturing both negative and positive correlations while building a data generating model and maintaining the positive-definiteness of the covariance matrix is challenging. Common positive-definite covariance structures, such as Compound Symmetry, AR(1), and Mat\'{e}rn, effectively handle positive correlations but have limited capacity for modeling negative correlations. Additionally, the observed connectivity matrix is not particularly sparse, making many known techniques for sparse covariance modeling unsuitable in this case \citep{MohammadiWit2015, Khondkeretal2013, Samantaetal2022}. 

Second, we observe clear patterns of connectivity in the marginal functional connectivity matrix that is lost when examining the conditional functional connectivity matrix (see Supplementary Material). This agrees with existing studies, which have shown that marginal functional connectivity tends to improve diagnostic performance compared to conditional connectivity \citep{ronicko2020diagnostic, haghighat2022sex}. 

Third, and most importantly, the voxel-level connectivity matrix shows a clear \emph{block} structure corresponding to the ROIs. This block structure arises from the hierarchical nature of brain parcellations, which nests voxels within a ROI, and ROIs within a network. Due to this structure, it is common in the literature to ``coarsen'' the outcomes by averaging the time series across voxels in each ROI at each time point, and to base the subsequent analysis on the ROI-level connectivity matrix. However, this method results in a substantial loss of information, as it fails to consider important details such as the size of each ROI and the variability among voxels in each ROI. Our replication efforts focus on making use of these rich high-resolution voxel-level data to jointly model participant-specific brain functional connectivity. 

Building on these observations, our aim is to develop a flexible model for the high-dimensional covariance matrix that can handle both positive and negative correlations, maintain positive-definiteness, and account for the matrix's inherent block structure. An important goal is to ensure that the model remains interpretable, allowing us to quantify how participant-specific covariates such as autism status influence the functional connectivity between voxels. To handle these requirements, we propose a block structure to model the high-dimensional covariance matrix. This approach has recently gained attention \citep[see, e.g.,][]{engle2012dynamic, archakov2022canonical, yang2023covariance, yang2023new} due to the widespread observation of organized block patterns in covariance matrices across various biomedical and financial data types. To the best of our knowledge, block structures have not yet been applied to extremely high-dimensional settings such as ours. Our block covariance structure specifies a parametric covariance model where each block shares the same covariance parameters, while allowing different parameters across blocks. This significantly reduces the number of parameters to be estimated while preserving the block structure. Among the various parameterization methods available for block structures, we adopt the canonical representation of block matrices proposed by \citet{archakov2022canonical}, which is particularly advantageous for efficient computation of key matrix quantities, such as the determinant and inverse. 

The above model, however, does not permit heterogeneity between participants due to participant-level characteristics. We further develop a block covariance model that incorporates the effect of participant-specific covariates $\boldsymbol{x}_{i} \in \mathbb{R}^{p}$. To achieve this, we carefully construct a covariate-dependent formulation for the Cholesky factors of the block covariance structure. This provides substantial flexibility by assigning different regression coefficients to each covariance block while ensuring the positive-definiteness of the covariance matrix. 
Due to their positive definiteness constraints, models that integrate covariates into the covariance matrix, otherwise referred to as \emph{covariance regression models}, often take forms that can be difficult to interpret. For example, \citet{chiu1996matrix, zhao2021covariate} model the elements of the logarithm of the covariance matrix as linear functions of covariates. \citet{zou2017covariance} link the covariance matrix to a linear combination of similarity or distance matrices of covariates. \citet{muschinski2022cholesky} explore distributional regression, linking parameters of the covariance matrix, derived through the modified Cholesky decomposition, to covariates. Models by \citet{hoff2012covariance, fox2015bayesian} result in a quadratic function of the covariates. This complexity often makes interpreting regression coefficients in covariance regression challenging.

While our model similarly produces a quadratic relationship between covariance and covariates, we show that our covariance model can be intuitively interpreted to estimate covariate effects on the \emph{dynamic} correlation. Dynamic correlation, a concept used in genomic analyses, explores how the correlation structure evolves in response to \emph{variations} in covariates. Unlike traditional methods that focus on linear or nonlinear first-order relationships, dynamic correlation emphasizes second-order patterns. For instance, \citet{li2002genome} introduced a statistical measure to quantify dynamic correlations between two genes modulated by a third gene. Building on this concept, subsequent studies have used dynamic correlation to model microarray expression data \citep{yu2018new,yang2022modeling}.

Finally, we construct a hypothesis testing framework to quantify the strength of association between ASD status and functional connectivity. Practically, we achieve this by employing a Bayesian hierarchical model that incorporates a continuous spike-and-slab prior \citep{george1993variable} for the ASD effect, which is expected to be sparse, alongside a continuous prior for other parameters of interest. Using Markov Chain Monte Carlo (MCMC) sampling, we calculate 95\% posterior credible intervals that quantify evidence of the influence of ASD on dynamic correlation. Given the Bayesian framework and the incorporation of covariates within a block covariance structure, we term our methodology the \textit{block covariance regression (BloCR)} model. As demonstrated in Section \ref{sec:Prior}, all model parameters can be efficiently sampled using a Gibbs sampler that depends only on low-dimensional summary statistics of the data, substantially improving memory usage and computational efficiency. 

The remainder of this paper is organized as follows. Section \ref{sec:method} introduces the BlocR model and explains how it enables an intuitive interpretation of covariate effects on dynamic correlation. Section \ref{sec:Prior} details the prior specifications and describes the Gibbs sampling procedure for efficient parameter estimation. Section \ref{sec:sim} presents simulation studies evaluating the model’s performance across different sparsity levels and data dimensions. Section \ref{sec:dat} investigates ASD-related dysconnections in ABIDE using our proposed BlocR model. Section \ref{sec:conc} concludes.

\section{Block covariance regression (BlocR) model} \label{sec:method}

\subsection{Canonical representation of block covariance matrices} 

In our analysis of ABIDE, let $i = 1,\dots,n$ represent the number of participants, and $t = 1, \dots, T_{i}$ represent the number of (thinned) time points for each participant $i$. The total number of rfMRI time points across all participants is given by $N = \sum_{i=1}^n T_{i}$. Let $M_{i}$ denote the number of voxels for participant $i$ in the voxel-level rfMRI data, with $M_{i} \approx 42{,}750$ in ABIDE. The $M_{i}$-dimensional vector of rfMRI outcomes for participant $i$ at time $t$ is denoted by $\boldsymbol{y}_{it}$, where the outcomes have been centered and scaled. We model the outcomes as independently following $M_{i}$-variate Gaussian distributions given by $\boldsymbol{y}_{it} \sim \mathcal{N}_{M_{i}} (\boldsymbol{0}, \boldsymbol{\Sigma}_{i})$. This assumes that the voxel-level covariance matrix $\boldsymbol{\Sigma}_{i} \in \mathbb{R}^{M_{i} \times M_{i}}$ is specific to each participant. 

A key feature of our modeling approach is the use of a block covariance structure for $\boldsymbol{\Sigma}_{i}$. This assumes a predetermined block structure within the covariance matrix, defining a parametric model where all elements within a block share the same covariance parameters, while allowing these parameters to vary across blocks. In ABIDE, the hierarchical organization of brain parcellations naturally induces this block structure. Let $J$ denote the number of blocks, and let the predetermined block partition be represented by $\boldsymbol{d}_{i} = (d_{i1}, \dots, d_{iJ})$, where $d_{ij}$ represents the size of the $j$-th block for $j = 1, \dots, J$ of the $i$-th participant, satisfying $M_{i} = \sum_{j=1}^{J} d_{ij}$. Since we use the same brain parcellation for all participants, the number of blocks $J$ remains constant across individuals. The number of voxels per participant may vary, leading to differences in block sizes across participants. Each covariance matrix $ \boldsymbol{\Sigma}_{i} $ can then be expressed in block form as follows: 
\begin{align}
\boldsymbol{\Sigma}_{i} = 
\begin{pmatrix}
\boldsymbol{\Sigma}_{i11} & \boldsymbol{\Sigma}_{i12} & \dots & \boldsymbol{\Sigma}_{i1J} \\  
\boldsymbol{\Sigma}_{i21} & \boldsymbol{\Sigma}_{i22} & \dots & \boldsymbol{\Sigma}_{i2J} \\  
\vdots & \vdots & \ddots & \vdots \\
\boldsymbol{\Sigma}_{iJ1} & \boldsymbol{\Sigma}_{iJ2} & \dots & \boldsymbol{\Sigma}_{iJJ} 
\end{pmatrix}. \label{eq:SigBlk}   
\end{align}
In equation \eqref{eq:SigBlk}, the $(j,\ell)$-th block is denoted as $\boldsymbol{\Sigma}_{ij\ell} \in \mathbb{R}^{d_{ij} \times d_{i\ell}}$. Define $\boldsymbol{1}_{d_{ij} \times d_{i\ell}} \in \mathbb{R}^{d_{ij} \times d_{i\ell}}$ the matrix of ones, $\boldsymbol{0}_{d_{ij} \times d_{i\ell}} \in \mathbb{R}^{d_{ij} \times d_{i\ell}}$ the matrix of zeros and $\boldsymbol{I}_{d_{ij}} \in \mathbb{R}^{d_{ij} \times d_{ij}}$ the identity matrix. To parameterize the block structure, we adopt the canonical representation proposed by \citet{archakov2022canonical}. Under this framework, each $(j,\ell)$-th block of $\boldsymbol{\Sigma}_{i}$ is defined as
\begin{equation} \label{eq:SigBlkEach}
\begin{split}
    \boldsymbol{\Sigma}_{ij\ell} &= \delta_{ij\ell} \boldsymbol{P}_{ij\ell} + \mathbbm{1}(j=\ell) \eta_{ij}\boldsymbol{P}_{ijj}^{\perp} ,
\end{split}
\end{equation}
where $\delta_{ij\ell} \in \mathbb{R}$ and $\eta_{ij} \in \mathbb{R}$ are block-specific parameters. The matrix $\boldsymbol{P}_{ij\ell} \in \mathbb{R}^{d_{ij} \times d_{i\ell}}$ is defined with all elements equal to $1/\sqrt{d_{ij}d_{i\ell}}$, i.e., $\boldsymbol{P}_{ij\ell} = 1/\sqrt{d_{ij}d_{i\ell}} \boldsymbol{1}_{d_{ij} \times d_{i\ell}}$. For diagonal blocks, the matrix $\boldsymbol{P}_{ijj}^{\perp} \in \mathbb{R}^{d_{ij} \times d_{ij}}$ is the complement of the projection matrix $\boldsymbol{P}_{ijj}$, defined as $\boldsymbol{P}_{ijj}^{\perp} = \boldsymbol{I}_{d_{ij}} -  1/d_{ij} \boldsymbol{1}_{d_{ij} \times d_{ij}}$. This formulation allows diagonal blocks to have different diagonal and off-diagonal elements, while all elements in an off-diagonal block are identical. 

Let $\boldsymbol{\nu}_{d_{ij}} = 1/\sqrt{d_{ij}} \boldsymbol{1}_{d_{ij} \times 1}$ denote a scaled vector of ones, and let $\boldsymbol{\nu}_{d_{ij}\perp} \in \mathbb{R}^{d_{ij} \times (d_{ij} - 1)}$ denote an orthonormal matrix orthogonal to $\boldsymbol{\nu}_{d_{ij}}$. For example, $\boldsymbol{\nu}_{d_{ij}\perp}$ can be constructed from the standard Helmert matrix \citep{lancaster1965helmert} of order $d_{ij}$, excluding its first row. Using these definitions, the matrices in equation \eqref{eq:SigBlkEach} can be expressed as $\boldsymbol{P}_{ij\ell} = \boldsymbol{\nu}_{d_{ij}} \boldsymbol{\nu}_{d_{i\ell}}^{\top}$ and $\boldsymbol{P}_{ijj}^{\perp} = \boldsymbol{\nu}_{d_{ij}\perp}\boldsymbol{\nu}_{d_{ij}\perp}^{\top}$. With these components, we construct the block matrices $\Tilde{\boldsymbol{\nu}}_{i} \in \mathbb{R}^{M_{i} \times J}$ and $\Tilde{\boldsymbol{\nu}}_{i\perp} \in \mathbb{R}^{M_{i} \times (M_{i}-J)}$ as follows: 
\begin{align*}
    \footnotesize
    \Tilde{\boldsymbol{\nu}}_{i} = \begin{pmatrix}
    \boldsymbol{\nu}_{d_{i1}} & \boldsymbol{0}_{d_{i1} \times 1} & \dots & \boldsymbol{0}_{d_{i1} \times 1} \\
    \boldsymbol{0}_{d_{i2} \times 1} & \boldsymbol{\nu}_{d_{i2}} & \dots & \boldsymbol{0}_{d_{i2} \times 1} \\
    \vdots & \vdots & \ddots & \vdots \\
    \boldsymbol{0}_{d_{iJ} \times 1} & \boldsymbol{0}_{d_{iJ} \times 1} & \dots & \boldsymbol{\nu}_{d_{iJ}} 
    \end{pmatrix}, \quad 
    \Tilde{\boldsymbol{\nu}}_{i\perp} = \begin{pmatrix}
    \boldsymbol{\nu}_{d_{i1}\perp} & \boldsymbol{0}_{d_{i1} \times (d_{i2}-1)} & \dots & \boldsymbol{0}_{d_{i1} \times (d_{iJ}-1)}\\
    \boldsymbol{0}_{d_{i2} \times (d_{i1}-1)} & \boldsymbol{\nu}_{d_{i2}\perp} & \dots & \boldsymbol{0}_{d_{i2} \times (d_{iJ}-1)}\\
    \vdots & \vdots & \ddots & \vdots \\
    \boldsymbol{0}_{d_{iJ} \times (d_{i1}-1)} & \boldsymbol{0}_{d_{iJ} \times (d_{i2}-1)} & \dots & \boldsymbol{\nu}_{d_{iJ}\perp}
    \end{pmatrix}. %\label{eq:Nus}
\end{align*}
Using $\Tilde{\boldsymbol{\nu}}_{i}$ and $\Tilde{\boldsymbol{\nu}}_{i\perp}$, we define the orthonormal matrix $\boldsymbol{Q}_{i}= (\Tilde{\boldsymbol{\nu}}_{i} \,\,\, \Tilde{\boldsymbol{\nu}}_{i\perp}) \in \mathbb{R}^{M_{i} \times M_{i}}$. With straightforward calculations, the covariance matrix $\boldsymbol{\Sigma}_{i}$ can then be written in its canonical form, 
\begin{align}
    \boldsymbol{\Sigma}_{i} = \boldsymbol{Q}_{i} \boldsymbol{D}_{i} \boldsymbol{Q}_{i}^{\top} \text{ where } \boldsymbol{D}_{i} = 
    \text{Block-diag}(\boldsymbol{\Delta}_{i}, \eta_{i1}\boldsymbol{I}_{d_{i1}-1}, \dots, \eta_{iJ}\boldsymbol{I}_{d_{iJ}-1}), \label{eq:SigCan}
\end{align}
and $\boldsymbol{\Delta}_{i} = (\delta_{ij\ell})_{j\ell = 1}^{J} \in \mathbb{R}^{J \times J}$. Here, $\boldsymbol{Q}_{i}$ serves to rotate $\boldsymbol{\Sigma}_{i}$ into its canonical form $\boldsymbol{D}_{i}$. This representation forms the basis of equation \eqref{eq:SigBlkEach}, connecting it to the canonical structure of $\boldsymbol{\Sigma}_{i}$. Equation \eqref{eq:SigCan} establishes that $\boldsymbol{\Sigma}_{i}$ is positive definite if and only if all $\eta_{ij} > 0$ and $\boldsymbol{\Delta}_{i}$ is positive definite. This insight is particularly valuable as it simplifies the task of constructing a structured, positive-definite high-dimensional covariance matrix $\boldsymbol{\Sigma}_{i}$. The parameterization is effectively reduced to two key components: $\boldsymbol{\Delta}_{i}$ and $\boldsymbol{\eta}_{i} = (\eta_{i1}, \dots, \eta_{iJ}) \in \mathbb{R}^{J}$, for $i = 1, \dots, n$.

\subsection{Covariate dependent modified Cholesky decomposition}

To guarantee that $\boldsymbol{\Sigma}_{i}$ is positive definite, it is necessary and sufficient that $\eta_{ij} > 0$ for all $j$ and that $\boldsymbol{\Delta}_{i}$ is positive definite. The first condition is straightforward to satisfy. Notably, the term $\eta_{ij} > 0$ influences the diagonal elements of the diagonal blocks $\boldsymbol{\Sigma}_{ijj}$, and can thus be interpreted as participant-specific, block-wise error terms.

To ensure that $\boldsymbol{\Delta}_{i}$ is positive definite, we parameterize it using the modified Cholesky decomposition, a method known for its mathematical and computational simplicity. This approach offers an unconstrained parameterization that guarantees positive definiteness without imposing complex constraints. The utility of the modified Cholesky decomposition has been highlighted in prior work on covariance matrix modeling \citep{muschinski2022cholesky, pourahmadi1999joint}. Our application is distinct in that we apply the decomposition specifically to $\boldsymbol{\Delta}_{i}$, rather than to the entire covariance matrix $\boldsymbol{\Sigma}_{i}$. A symmetric matrix $\boldsymbol{\Delta}_{i}$ is positive definite if and only if there exists a unique unit lower triangular matrix $\boldsymbol{L}_{i}$ and a unique diagonal matrix $\boldsymbol{\Lambda}_{i}$ with positive diagonal entries such that $\boldsymbol{\Delta}_{i} = \boldsymbol{L}_{i}\boldsymbol{\Lambda}_{i}\boldsymbol{L}_{i}^{\top}$.
In this formulation, $\boldsymbol{L}_{i}$ is straightforward to model, as its lower off-diagonal entries can take any real value without constraint. Positive definiteness is ensured by requiring that the diagonal entries of $\boldsymbol{\Lambda}_{i}$, denoted as $\boldsymbol{\lambda}_{i} = (\lambda_{i1}, \dots, \lambda_{ij}) \in \mathbb{R}^{J}$, satisfy $\lambda_{ij} > 0$ for all $j = 1, \dots, J$.

Our primary objective is to examine the effect of participant-specific covariates on functional connectivity between voxels. To achieve this, we propose introducing a linear model of covariates $\boldsymbol{x}_{i} = (x_{i1}, \dots, x_{ip}) \in \mathbb{R}^{p}$ to each element of the Cholesky factors. Specifically, we define the entries of the unit lower triangular matrix $\boldsymbol{L}_{i}$ as follows:
\begin{align*}
    &L_{i\ell\ell} = 1; \quad L_{ij\ell} = 0 \text{ for } \ell > j; \quad L_{ij\ell} = \boldsymbol{x}_{i}^{\top} \boldsymbol{\beta}_{j\ell} \text{ for } \ell < j, \; j = 2, \dots, J,
\end{align*}
where the lower triangular off-diagonal elements $L_{ij\ell}$ are modeled as linear functions of the covariates, allowing full flexibility by assigning distinct regression coefficients $\boldsymbol{\beta}_{j\ell} = (\beta_{1j\ell}, \dots, \beta_{pj\ell}) \in \mathbb{R}^{p}$ to each $L_{ij\ell}$. This setup leads to the following expression for the entries of $\boldsymbol{\Delta}_{i} = (\delta_{ij\ell})_{j\ell = 1}^{J}$, and consequently for $\boldsymbol{\Sigma}_{ij\ell}$: 
\begin{align}
    \delta_{ij\ell} &= \sum_{\ell'=1}^{\ell} \lambda_{i,\ell'} L_{ij\ell'} L_{i\ell\ell'} = \sum_{\ell'=1}^{\ell} \lambda_{i\ell'} \boldsymbol{x}_{i}^{\top} \boldsymbol{\beta}_{j\ell'}\boldsymbol{\beta}_{\ell\ell'}^{\top} \boldsymbol{x}_{i}. \label{eq:Delta}
\end{align}
Substituting equation \eqref{eq:Delta} into equation \eqref{eq:SigBlkEach}, the covariance block $\boldsymbol{\Sigma}_{ij\ell}$ is expressed as:
\begin{align}
    \boldsymbol{\Sigma}_{ij\ell} = \frac{1}{\sqrt{d_{ij}d_{i\ell}}} \Bigg(\sum_{\ell'=1}^{\ell} \lambda_{i\ell'} \boldsymbol{x}_{i}^{\top} \boldsymbol{\beta}_{j\ell'}\boldsymbol{\beta}_{\ell\ell'}^{\top} \boldsymbol{x}_{i} \Bigg) \boldsymbol{1}_{d_{ij} \times d_{i\ell}} + \mathbbm{1}(j=\ell) \eta_{ij}\boldsymbol{P}_{ijj}^{\perp}. \label{eq:Sigma}
\end{align}
This formulation shows that the covariance block is a quadratic function of the covariates, scaled by the Cholesky diagonal entries $\lambda_{i\ell'}$ and the block sizes $d_{ij}$ and $d_{i\ell}$. The flexibility of $\boldsymbol{\Sigma}_{ij\ell}$ stems from the assignment of distinct regression coefficients $\boldsymbol{\beta}_{j\ell}$ to each off-diagonal element $L_{ij\ell}$, allowing the covariance structure to adaptively capture the effects of covariates for each participant and each block. 

\subsection{Interpreting covariate effects via second-order patterns}

The representation of the covariance block $\boldsymbol{\Sigma}_{ij\ell}$ in equation \eqref{eq:Sigma}, and consequently the entire covariance matrix, appears fairly complicated, a common characteristic of covariance regression models. These models must satisfy positive definiteness constraints, often resulting in intricate forms that are challenging to interpret. This issue has been noted in various modeling strategies in the literature \citep{chiu1996matrix, zhao2021covariate, zou2017covariance, muschinski2022cholesky, hoff2012covariance, fox2015bayesian} and applies to our model as well. The inherent complexity of the model can obscure the interpretation of regression coefficients in covariance regression models. In our approach, however, we demonstrate that the seemingly complex structure of the covariance matrix provides a clear and interpretable explanation of the covariate effects. 

Rather than directly interpreting the quadratic first-order patterns, we focus on second-order patterns. This approach aligns with the interpretation of dynamic correlation \citep{li2002genome, yu2018new, yang2022modeling}, a framework used in genomic analyses to explore how correlation structures between genetic markers evolve in response to variations in covariates. %Emphasizing second-order patterns rather than traditional first-order relationships can offer a more comprehensive understanding of the complex structure of the covariance matrix. 
In our analysis, we investigate how the functional connectivity changes in response to variations in participant-specific clinical characteristics. In ABIDE, these clinical variables include ASD diagnostic group, age at the time of the scan, sex, group-sex interaction, and eye status during the resting scan. Additional details about these clinical variables and the rationale behind their selection are provided in Section \ref{sec:dat}. 

To quantify this relationship, we take the partial derivative of the covariance block $\boldsymbol{\Sigma}_{ij\ell}$ with respect to the $q$th covariate $x_{iq}$: 
\begin{align}
    \frac{\partial \boldsymbol{\Sigma}_{ij\ell}}{\partial x_{iq}} 
   = \Bigg\{ \frac{1}{\sqrt{d_{ij}d_{i\ell}}}  \sum_{\ell'=1}^{\ell} \lambda_{i\ell'} \Big( \beta_{qj\ell'} \boldsymbol{\beta}_{\ell\ell'}^{\top}\boldsymbol{x}_{i} + \beta_{p\ell\ell'} \boldsymbol{\beta}_{j\ell'}^{\top}\boldsymbol{x}_{i} \Big)  \Bigg\} \boldsymbol{1}_{d_{j} \times d_{\ell}}. \label{eq:deriv}
\end{align}
The partial derivative $\partial \boldsymbol{\Sigma}_{ij\ell}/\partial x_{iq}$ in equation \eqref{eq:deriv} measures the sensitivity of covariance between voxels in the $(j,\ell)$-th ROI to changes in the $q$-th participant-specific covariate $x_{iq}$. When the covariate is binary, a similar expression can be derived using a finite difference approach. The corresponding formulation is provided in Section \ref{sec:dat} for the analysis of ABIDE, where the covariate of interest (ASD diagnostic group) is binary. 

Large absolute values of these derivatives indicate strong evidence of dynamic correlation, meaning that the covariance between voxels in the $(j,\ell)$-th ROI are highly influenced by changes in $x_{iq}$. Conversely, values close to zero suggest weak or no dynamic correlation, indicating minimal influence of changes in $x_{iq}$ on the covariance between voxels in the ROI. This analysis corresponds to a hypothesis test of the following form: 
\begin{itemize}[leftmargin=*]
    \item $H_{0}$: Changes in $x_{iq}$ have no association with the values in the covariance block $\boldsymbol{\Sigma}_{ij\ell}$ as described by the model.  
    \item $H_{1}$: Changes in $x_{iq}$ are associated with the values in the covarince block $\boldsymbol{\Sigma}_{ij\ell}$ as described by the model.  
\end{itemize}
To conduct this hypothesis test, we employ a Bayesian hierarchical modeling approach. Using Markov chain Monte Carlo (MCMC) samples from the posterior distributions of parameters of interest, we calculate 95\% posterior credible intervals for equation \eqref{eq:deriv}. If the credible interval does not include zero, this provides strong evidence of dynamic correlation, leading to the rejection of the null hypothesis. To ensure the model can effectively capture scenarios where $\partial \boldsymbol{\Sigma}_{ij\ell}/\partial x_{iq} = 0$, we employ a continuous spike-and-slab prior \citep{george1993variable} on the regression coefficients where sparsity is expected, alongside a continuous prior for other parameters. This approach enhances the model's ability to discern cases of no dynamic correlation while maintaining its flexibility to detect strong evidence when present.

\section{Prior specifications and Gibbs sampler} \label{sec:Prior}

Given the complexity of our BlocR model and the scale of the data, an efficient computational approach is essential. This section describes the prior distributions of the model parameters and demonstrates how conditional posterior distributions can be efficiently sampled using the Gibbs sampler. A key component of this approach is leveraging algebraic simplifications to construct a Gibbs sampling scheme that depends only on low-dimensional summaries of the data. We first present the complete hierarchical model formulation for each participant $i=1,\dots,n$: 
\begin{equation}
    \begin{aligned}
    &\eta_{ij} \sim \text{Inv-Gamma}(a_{0},b_{0}) \text{ for } j=1,\dots,J,\\
    &\lambda_{ij} \sim \text{Inv-Gamma}(a_{1},b_{1}) \text{ for } j=1,\dots,J,\\
    &\pi_{j\ell} \sim \text{Ber}(q_{1}) \text{ for } \ell < j, \, j=2,\dots,J,\\
    &\beta_{pj\ell} \,|\, \pi_{j\ell} \sim (1-\pi_{j\ell}) \mathcal{N}(0,\tau_{0}^{2})  + \pi_{j\ell} \mathcal{N}(0,\tau_{1}^{2}) \text{ for } \ell < j, \, j=2,\dots,J,\\
    &\beta_{qj\ell} \sim \mathcal{N}(0,\tau_{2}^{2}) \text{ for } q=1,\dots,p-1 \text{ and } \ell < j, \, j=2,\dots,J, \\
    &L_{ij\ell} = \boldsymbol{x}_{i}^{\top} \boldsymbol{\beta}_{j\ell} \text{ for } \ell < j, \, j=2,\dots,J,\\
    &\boldsymbol{\Delta}_{i} = \boldsymbol{L}_{i}\boldsymbol{\Lambda}_{i}\boldsymbol{L}_{i}^{\top}, \\
    &\boldsymbol{D}_{i} = 
    \text{Block-diag}(\boldsymbol{\Delta}_{i}, \eta_{i1}\boldsymbol{I}_{d_{1}-1}, \dots, \eta_{ij}\boldsymbol{I}_{d_{J}-1}), \\
    &\boldsymbol{\Sigma}_{i} = \boldsymbol{Q} \boldsymbol{D}_{i} \boldsymbol{Q}^{\top},\\
    &\boldsymbol{y}_{it} \,|\, \boldsymbol{\eta}_{i}, \boldsymbol{\lambda}_{i}, \boldsymbol{\beta} \sim \mathcal{N}_{M_{i}}(\boldsymbol{0}, \boldsymbol{\Sigma}_{i}) \text{ for }t=1,\dots,T_{i}. \label{eq:Model}
    \end{aligned}
\end{equation}

\subsection{Simplified likelihood formulation for Gibbs sampling}

For each $M_{i}$-variate outcome $\boldsymbol{y}_{it} \in \mathbb{R}^{M_{i}}$, we define $\boldsymbol{Y}_{i} = (\boldsymbol{y}_{i1}, \dots, \boldsymbol{y}_{iT_{i}})^{\top} \in \mathbb{R}^{T_{i} \times M_{i}}$. The sample covariance for the $i$-th participant is given by $\boldsymbol{S}_{i} = 1/T_{i} \sum_{t=1}^{T_{i}} \boldsymbol{y}_{it}\boldsymbol{y}_{it}^{\top} = \boldsymbol{Y}_{i}^{\top} \boldsymbol{Y}_{i} / T_{i} \in \mathbb{R}^{M_i \times M_i}$. We define $\boldsymbol{A}_{i} = \Tilde{\boldsymbol{\nu}}_{i}^{\top}\boldsymbol{S}_{i}\Tilde{\boldsymbol{\nu}}_{i} \in \mathbb{R}^{J \times J}$, where its entries, $(a_{ij\ell})_{j,\ell=1}^{J}$, are given by $a_{ij\ell} = \boldsymbol{\nu}_{ij}^{\top}\boldsymbol{S}_{ij\ell}\boldsymbol{\nu}_{i\ell} = 1/\sqrt{d_{ij}d_{i\ell}} \boldsymbol{1}_{d_{ij}}^{\top}\boldsymbol{S}_{ij\ell}\boldsymbol{1}_{d_{i\ell}}$. Here, $\boldsymbol{S}_{ij\ell} \in \mathbb{R}^{d_{ij} \times d_{i\ell}}$ represents the $(j,\ell)$-th block of $\boldsymbol{S}_{i}$, and $\boldsymbol{A}_{i,1:j, 1:j} \in \mathbb{R}^{j \times j}$ represents the upper-left submatrix of $\boldsymbol{A}_{i}$, containing the first $j$ rows and columns.  

We introduce the unit upper triangular matrix $\boldsymbol{U}_{i} = (\boldsymbol{L}_{i}^{-1})^{\top} \in \mathbb{R}^{J \times J}$. Since the inverse of a lower triangular matrix is also lower triangular, $\boldsymbol{U}_i$ follows naturally from this property. The vector $\boldsymbol{U}_{i,1:j,j} = (U_{i1j}, U_{i2j}, \dots, U_{ijj}) \in \mathbb{R}^{j}$ consists of the first $j$ entries of the $j$-th column of $\boldsymbol{U}_{i}$, with $U_{ijj} = 1$ by construction.

We first present our main result, with a proof given below.

\begin{proposition}[Simplified likelihood function] \label{prop:likelihood}
The likelihood corresponding to the model specified in (\ref{eq:Model}) simplifies to  
\begin{equation}
\begin{split}
&\prod_{i=1}^{n} p(\boldsymbol{Y}_{i} \,|\, \boldsymbol{\eta}_{i}, \boldsymbol{\lambda}_{i}, \boldsymbol{\beta}) \propto \\
&\Big\{ \prod_{i=1}^{n}\prod_{j=1}^{J} \eta_{ij}^{-\frac{T_{i}(d_{ij}-1)}{2}} \Big\} \exp\Big[-\frac{1}{2} \sum_{i=1}^{n}\sum_{j=1}^{J} T_{i}\eta_{ij}^{-1} \Big\{ \operatorname{tr}(\boldsymbol{S}_{ijj}) - \frac{1}{d_{ij}}\boldsymbol{1}_{d_{ij}}^{\top}\boldsymbol{S}_{ijj}\boldsymbol{1}_{d_{ij}}   \Big\} \Big] \\
&\Big(\prod_{i=1}^{n} \prod_{j=1}^{J} \lambda_{ij}^{-\frac{T_{i}}{2}} \Big) \exp\Big(-\frac{1}{2} \sum_{i=1}^{n}\sum_{j=1}^{J} T_{i} \lambda_{ij}^{-1} \boldsymbol{U}_{i,1:j,j}^{\top} \boldsymbol{A}_{i,1:j,1:j} \boldsymbol{U}_{i,1:j,j} \Big). \label{eq:Like}
\end{split}
\end{equation}
\end{proposition}
A key advantage of this expression is that all $i=1,\dots,n$ and $j=1,\dots,J$ components can be decomposed into products, leading to a computationally efficient form used throughout the derivation of conditional posterior distributions. 

Proposition \ref{prop:likelihood} allows for substantial computational and privacy advantages. The model structure allows the simplified likelihood in equation \eqref{eq:Like} to be computed without requiring participant-level outcomes. Instead, it relies on participant-level summary statistics derived from sample covariance matrices: $\operatorname{tr}(\boldsymbol{S}_{ijj}), \boldsymbol{1}_{d_{ij}}^{\top}\boldsymbol{S}_{ijj}\boldsymbol{1}_{d_{ij}}, \boldsymbol{A}_{i,1:j,1:j}$ for $j=1,\dots,J$. These summary statistics need only be calculated once before beginning MCMC sampling, providing huge computational and memory gains. This approach ensures computational feasibility for the massive ABIDE data, which consist over one trillion data points. Additionally, in other applications that may require access to sensitive data, our formulation guarantees data privacy by avoiding direct use of participant-level observations, relying instead on aggregated summary measures.

Further, Proposition \ref{prop:likelihood} suggests a natural estimator for $\boldsymbol{\Delta}_{i}$. Indeed, by construction, the matrix $\boldsymbol{A}_{i}$ serves as a reasonable estimate of $\boldsymbol{\Delta}_{i}$. This property will be useful in Sections \ref{sec:sim} and \ref{sec:dat}, where we evaluate the accuracy of our model by comparing $\boldsymbol{\Delta}_{i}$ calculated from maximum a posteriori (MAP) estimates of the parameters with the observed $\boldsymbol{A}_{i}$.

The derivations supporting these key results are presented in the following proof. 
\begin{proof}
The likelihood for the $i$-th participant is expressed as:  
\begin{align}
    p(\boldsymbol{Y}_{i} \,|\, \boldsymbol{\eta}_{i}, \boldsymbol{\lambda}_{i}, \boldsymbol{\beta}) &\propto \text{det}(\boldsymbol{\Sigma}_{i})^{-T_{i}/2} \exp\Big\{-\frac{T_{i}}{2} \operatorname{tr}(\boldsymbol{S}_{i}\boldsymbol{\Sigma}_{i}^{-1})\Big\}. \label{eq:like}
\end{align}
To simplify this expression, we examine the participant components of equation \eqref{eq:like}. Using the canonical representation of $\boldsymbol{\Sigma}_{i}$ from equation \eqref{eq:SigCan} and noting that $\boldsymbol{Q}_{i}$ is an orthonormal matrix, the determinant in equation \eqref{eq:like} can be rewritten as:
\begin{align}
    \det(\boldsymbol{\Sigma}_{i}) = \det(\boldsymbol{Q}_{i} \boldsymbol{D}_{i} \boldsymbol{Q}^{\top}_{i}) = \det(\boldsymbol{D}_{i}) = \prod_{j=1}^{J} \lambda_{ij} \eta_{ij}^{d_{ij}-1}. \label{eq:detSig}
\end{align}
Similarly, the precision matrix $\boldsymbol{\Sigma}_{i}^{-1}$ can also be derived using the canonical representation of $\boldsymbol{\Sigma}_{i}$:
\begin{equation}\label{eq:Prec}
\begin{split}
\boldsymbol{\Sigma}_{i}^{-1} 
&= \boldsymbol{Q}_{i} \boldsymbol{D}_{i}^{-1} \boldsymbol{Q}_{i}^{\top} = \Tilde{\boldsymbol{\nu}}_{i}\boldsymbol{\Delta}_{i}^{-1}\Tilde{\boldsymbol{\nu}}_{i}^{\top} + \text{Block-diag}(\eta_{i1}^{-1} \boldsymbol{P}_{i11}^{\perp}, \dots, \eta_{ij}^{-1} \boldsymbol{P}_{ijj}^{\perp}).
\end{split}
\end{equation}
Substituting  equation \eqref{eq:Prec} into the trace term in equation \eqref{eq:like}, we obtain:
\begin{equation} \label{eq:trace}
\begin{split}
\operatorname{tr}(\boldsymbol{S}_{i}\boldsymbol{\Sigma}_{i}^{-1})
&= \operatorname{tr}(\boldsymbol{S}_{i}\Tilde{\boldsymbol{\nu}}\boldsymbol{\Delta}_{i}^{-1}\Tilde{\boldsymbol{\nu}}^{\top}) + \textstyle \sum_{j=1}^{J} \eta_{ij}^{-1} \Big\{ \operatorname{tr}(\boldsymbol{S}_{ijj}) - \frac{1}{d_{ij}}\boldsymbol{1}_{d_{ij}}^{\top}\boldsymbol{S}_{ijj}\boldsymbol{1}_{d_{ij}}   \Big\},
\end{split}
\end{equation}
Using the modified Cholesky decomposition of $\boldsymbol{\Delta}_{i}$, we express $\boldsymbol{\Delta}_{i}^{-1}$ as: 
\begin{align}
    &\boldsymbol{\Delta}_{i}^{-1} = (\boldsymbol{L}_{i}^{-1})^{\top} \boldsymbol{\Lambda}_{i}^{-1} \boldsymbol{L}_{i}^{-1} = \boldsymbol{U}_{i} \boldsymbol{\Lambda}_{i}^{-1} \boldsymbol{U}_{i}^{\top}. \label{eq:Ui}
\end{align}
Substituting equation \eqref{eq:Ui} into the first term in equation \eqref{eq:trace}, we can simplify the term involving $\boldsymbol{\Delta}_{i}^{-1}$ as follows: 
\begin{align}
\operatorname{tr}(\boldsymbol{S}_{i}\Tilde{\boldsymbol{\nu}}_{i}\boldsymbol{\Delta}_{i}^{-1}\Tilde{\boldsymbol{\nu}}_{i}^{\top}) &= \operatorname{tr}(\Tilde{\boldsymbol{\nu}}_{i}^{\top}\boldsymbol{S}_{i}\Tilde{\boldsymbol{\nu}}_{i} \, \boldsymbol{U}_{i} \boldsymbol{\Lambda}_{i}^{-1} \boldsymbol{U}_{i}^{\top}) = \operatorname{tr}(\boldsymbol{\Lambda}_{i}^{-1}\boldsymbol{U}_{i}^{\top}\boldsymbol{A}_{i}\boldsymbol{U}_{i}), \label{eq:trSA}
\end{align} 
To calculate the diagonal components of the term in equation \eqref{eq:trSA}, we write: 
\begin{align}(\boldsymbol{\Lambda}_{i}^{-1}\boldsymbol{U}_{i}^{\top}\boldsymbol{A}_{i}\boldsymbol{U}_{i})_{jj} 
&= \lambda_{ij}^{-1} \boldsymbol{U}_{i,1:j,j}^{\top} \boldsymbol{A}_{i,1:j,1:j} \boldsymbol{U}_{i,1:j,j} \text{ for } j=1,\dots,J, \label{eq:UAU}
\end{align}
By combining Equations \eqref{eq:detSig}, \eqref{eq:trace}, \eqref{eq:trSA}, and \eqref{eq:UAU}, we obtain the simplified likelihood form given in Equation \eqref{eq:Like}. 
\end{proof}

\subsection{Gibbs sampling: Full conditional distributions of  \texorpdfstring{$\eta_{ij}$}{Lg} and \texorpdfstring{$\lambda_{ij}$}{Lg}}

Independent priors are assigned to the block-wise error terms $\boldsymbol{\eta} = (\boldsymbol{\eta}_{1}, \dots, \boldsymbol{\eta}_{n}) \in \mathbb{R}^{nJ}$, with $\eta_{ij} \sim \text{Inv-Gamma}(a_{0},b_{0})$. Similarly, independent priors are assigned to the Cholesky diagonal entries $\boldsymbol{\lambda} = (\boldsymbol{\lambda}_{1}, \dots, \boldsymbol{\lambda}_{n}) \in \mathbb{R}^{nJ}$, with $\lambda_{ij} \sim \text{Inv-Gamma}(a_{1},b_{1})$. 

Leveraging the conjugacy of the Gaussian likelihood in equation \eqref{eq:Like} and the Inverse-Gamma prior, we can derive the conditional posterior distribution for $\eta_{ij}$ and $\lambda_{ij}$ in closed form. Using the Gibbs sampler, we update $\eta_{ij}$ from the following conditionally independent posteriors: 
\begin{align*}
    p(\eta_{ij} \,|\, \cdot) \propto \text{Inv-Gamma} \Big[a_{0} + \frac{T_{i}(d_{ij}-1)}{2}, b_{0} + \frac{T_{i}}{2} \Big\{ tr(\boldsymbol{S}_{jj}) - \frac{1}{d_{ij}}\boldsymbol{1}_{d_{ij}}^{\top}\boldsymbol{S}_{ijj}\boldsymbol{1}_{d_{ij}}   \Big\} \Big].
\end{align*}
Similarly, we sample $\lambda_{ij}$ from the following conditionally independent posteriors: 
\begin{align*}
    p(\lambda_{ij} \,|\, \cdot) \propto \text{Inv-Gamma} \Big(a_{1} + \frac{T_{i}}{2}, b_{1} + \frac{T_{i}}{2} \boldsymbol{U}_{i,1:j,j}^{\top} \boldsymbol{A}_{i,1:j,1:j} \boldsymbol{U}_{i,1:j,j} \Big).
\end{align*}

\subsection{Gibbs sampling: Full conditional distribution of \texorpdfstring{$\boldsymbol{\beta}_{j}$}{Lg}}

This subsection describes the efficient Gibbs sampling procedure for the regression coefficient parameters, first outlining key results followed by their derivations. To facilitate our discussion, we express the regression coefficients in a vectorized format: 
\begin{equation}
\begin{split}
&\boldsymbol{\beta}_{q,j,1:(j-1)} = (\beta_{qj1}, \dots, \beta_{qj,j-1}) \in \mathbb{R}^{j-1} \text{ for } q = 1,\dots,p, \\
&\boldsymbol{\beta}_{j} := \boldsymbol{\beta}_{1:p,j,1:(j-1)} = (\boldsymbol{\beta}_{1,j,1:(j-1)}, \dots, \boldsymbol{\beta}_{p,j,1:(j-1)}) \in \mathbb{R}^{p(j-1)} \text{ for } j=2,\dots,J\\ \label{eq:Betas}
&\boldsymbol{\beta} = (\boldsymbol{\beta}_{2}, \dots, \boldsymbol{\beta}_{J}) \in \mathbb{R}^{pJ(J-1)/2}.
\end{split}
\end{equation}

Among the $p$ covariates, we apply a continuous spike-and-slab prior \citep{george1993variable} exclusively to the regression coefficient parameters associated with the $p$-th covariate, $\boldsymbol{\beta}_{p,j,1:(j-1)}$. For the regression coefficients corresponding to the remaining covariates, $\boldsymbol{\beta}_{q,j,1:(j-1)}$ for $q=1,\dots,p-1$, we use Gaussian priors. This prioritization reflects the ABIDE analysis in Section \ref{sec:dat}, where the ASD diagnostic group membership variable is the primary focus for interpretation among the regression coefficients. The induced sparsity in $\boldsymbol{\beta}_{p,j,1:(j-1)}$ helps capture scenarios where $\partial \boldsymbol{\Sigma}_{ij\ell}/\partial x_{ip} = 0$. Naturally, this framework can be extended to incorporate sparsity in the regression coefficients of other covariates as needed, without loss of generality. The priors are specified as follows:
\begin{equation} \label{eq:AllBetaPriors}
    \begin{split}
    &\pi_{j\ell} \sim \text{Ber}(q_{1}) \text{ for } \ell < j, \, j=2,\dots,J, \\
    &(\beta_{pj\ell} \,|\, \pi_{j\ell}) \sim (1-\pi_{j\ell}) \mathcal{N}(0, \tau_{0}^{2}) + \pi_{j\ell} \mathcal{N}(0, \tau_{1}^{2}), \\
    &\boldsymbol{\beta}_{q,j,1:(j-1)} \sim \mathcal{N}(\boldsymbol{0}, \tau_{2}^{2}\boldsymbol{I}_{j-1}) \text{ for } q=1,\dots,p-1.
    \end{split}
\end{equation}
Here, $\mathcal{N}(0,\tau_{1}^{2})$ and $\mathcal{N}(0,\tau_{0}^{2})$ represent the spike and slab components of the prior, respectively, with hyperparameters $\tau_{2}^{2} > 0$ and $\tau_{1}^{2} \gg \tau_{0}^{2} > 0$. The mixing probabilities $\pi_{j\ell}$ determine whether a coefficient is drawn from the slab or spike distribution. In high-dimensional settings, the mixing probability hyperparameter $q_{1}$ is often chosen to be small. In our simulation experiments and ABIDE analysis, we set $q_{1}=0.5$ by default. The use of a continuous spike-and-slab prior allows for effective modeling of sparsity, as these priors have become a cornerstone of Bayesian variable selection \citep{tadesse2021handbook, biswas2022scalable}. 

\begin{proposition}[Full Conditional Posterior distribution of $\boldsymbol{\beta}_{j}$] \label{prop:betas}
The conditional posterior distribution of  $\boldsymbol{\beta}_{j}$ corresponding to the prior specified in (\ref{eq:AllBetaPriors}) and the hierarchical model in (\ref{eq:Model}) can be expressed as
\begin{align}
   &\boldsymbol{\beta}_{j} = \boldsymbol{C}_{j}^{-1/2} \Big\{\boldsymbol{C}_{j}^{-1/2} \boldsymbol{\mu}_{j} + \boldsymbol{Z} \Big\} \text{ where } \boldsymbol{Z} \sim \mathcal{N}(\boldsymbol{0}, \boldsymbol{I}_{p(j-1)}), \label{eq:SampBeta}
\end{align}
with $\boldsymbol{\mu}_{j} \in \mathbb{R}^{p(j-1)}$ and $\boldsymbol{C}_{j} \in \mathbb{R}^{p(j-1) \times p(j-1)}$ given by 
\begin{align*}
    \boldsymbol{\mu}_{j} &= \sum_{i=1}^{n} T_{i} \lambda_{ij}^{-1} \Tilde{\boldsymbol{x}}_{i} \mathbf{U}_{i,1:(j-1),1:(j-1)}^{\top} \boldsymbol{A}_{i,1:(j-1),j}  \\
    \boldsymbol{C}_{j} &= \sum_{i=1}^{n} T_{i}\lambda_{ij}^{-1} \Tilde{\boldsymbol{x}}_{i} \mathbf{U}_{i,1:(j-1),1:(j-1)}^{\top} \boldsymbol{A}_{i,1:(j-1),1:(j-1)} \mathbf{U}_{i,1:(j-1),1:(j-1)} \Tilde{\boldsymbol{x}}_{i}^{\top} + \boldsymbol{\Pi}_{j}.
\end{align*}
\end{proposition}

In Proposition \ref{prop:betas}, $\Tilde{\boldsymbol{x}}_{i} = (\boldsymbol{x}_{i} \otimes \boldsymbol{I}_{j-1}) \in \mathbb{R}^{p(j-1) \times (j-1)}$. The diagonal matrix $\boldsymbol{\Pi}_{j} \in \mathbb{R}^{p(j-1) \times p(j-1)}$ has the following structure: the first $(p-1)(j-1)$ diagonal elements are set to $\tau_{2}^{-2}$. For the remaining $(j-1)$ diagonal elements, the value of the $\ell$-th term depends on $\pi_{j\ell}$. If $\pi_{j\ell} = 1$, the diagonal term is $\tau_{1}^{-2}$ (spike); if $\pi_{j\ell} = 0$, the diagonal term is $\tau_{0}^{-2}$ (slab). This formulation provides an explicit Gibbs sampling scheme for $\boldsymbol{\beta}_{j}$, with the inverse term $\boldsymbol{C}_{j}^{-1/2}$ computed via eigendecomposition.

The derivation of the conditional posterior distribution of $\boldsymbol{\beta}_{j}$ in Proposition \ref{prop:betas} is given in the following proof. 

\begin{proof}
We first recall the vectorized representation in equation \eqref{eq:Betas}, which facilitates prior specification. This representation differs from an alternative format, $\boldsymbol{\beta}_{j\ell} = (\beta_{1j\ell}, \dots, \beta_{pj\ell}) \in \mathbb{R}^{p}$ for $\ell < j$, $j=2,\dots,J$, which is primarily used for modeling the covariance block $\boldsymbol{\Sigma}_{ij\ell}$ and interpreting second-order patterns.

Let $\boldsymbol{\pi}_{j} := \boldsymbol{\pi}_{j,1:(j-1)} = (\pi_{j1}, \dots, \pi_{j,j-1}) \in \mathbb{R}^{j-1}$ denote the vector of mixing probabilities. For $j = 2,\dots,J$, the priors on the regression coefficients can be expressed in vectorized form as follows:
\begin{align}
    &p(\boldsymbol{\beta}_{j} \,|\, \boldsymbol{\pi}_{j}) \propto p(\boldsymbol{\beta}_{1,j,1:(j-1)},\dots,\boldsymbol{\beta}_{p-1,j,1:(j-1)}) \, p(\boldsymbol{\beta}_{p,j,1:(j-1)} \,|\, \pi_{\ell,j}) = \mathcal{N}(\boldsymbol{0}, \boldsymbol{\Pi}_{j}^{-1}). \label{eq:BetaPrior}
\end{align}
A key step in deriving the conditional posterior distribution is obtaining a closed-form expression for the unit upper triangular matrix $\boldsymbol{U}_{i} = (\boldsymbol{L}_{i}^{-1})^{\top}$. Solving the equation $\boldsymbol{U}_{i}\boldsymbol{U}_{i}^{-1} = \boldsymbol{U}_{i}\boldsymbol{L}_{i}^{\top} = \boldsymbol{I}_{J}$ via forward substitution, for each $\ell < j$ and $j = 2,\dots,J$, we obtain:
\begin{align*}
    &U_{i\ell j} = -L_{ij\ell} - \sum_{\ell'=\ell+1}^{j-1} U_{i\ell\ell'}L_{ij\ell'} = - \boldsymbol{x}_{i}^{\top}\boldsymbol{\beta}_{j\ell} - \sum_{\ell'=\ell+1}^{j-1} U_{i\ell\ell'} \Big(\boldsymbol{x}_{i}^{\top}\boldsymbol{\beta}_{j\ell'} \Big).
\end{align*}
From this, the $j$-th subcolumn $\boldsymbol{U}_{i,1:(j-1),j} \in \mathbb{R}^{j-1}$ can be written as:
\begin{align}
    \boldsymbol{U}_{i,1:(j-1),j} 
    &= - \mathbf{U}_{i,1:(j-1),1:(j-1)} \Tilde{\boldsymbol{x}}_{i}^{\top} \boldsymbol{\beta}_{j}, \label{eq:ForSub}
\end{align}
Substituting equation \eqref{eq:ForSub} into the likelihood term from equation \eqref{eq:Like}, we rewrite:
\begin{equation} \label{eq:UtAU}
\begin{split}
&\boldsymbol{U}_{i,1:j,j}^{\top} \boldsymbol{A}_{i,1:j,1:j} \boldsymbol{U}_{i,1:j,j} \\
&= a_{ijj} + 2\boldsymbol{A}_{i,j,1:(j-1)}\boldsymbol{U}_{i,1:(j-1),j} + \boldsymbol{U}_{i,1:(j-1),j}^{\top}\boldsymbol{A}_{i,1:(j-1),1:(j-1)}\boldsymbol{U}_{i,1:(j-1),j} \\
&= a_{ijj} - 2\boldsymbol{A}_{i,j,1:(j-1)}\mathbf{U}_{i,1:(j-1),1:(j-1)} \Tilde{\boldsymbol{x}}_{i}^{\top} \boldsymbol{\beta}_{j} \\
&\quad + \boldsymbol{\beta}_{j}^{\top}\Tilde{\boldsymbol{x}}_{i} \mathbf{U}_{i,1:(j-1),1:(j-1)}^{\top} \boldsymbol{A}_{i,1:(j-1),1:(j-1)} \mathbf{U}_{i,1:(j-1),1:(j-1)} \Tilde{\boldsymbol{x}}_{i}^{\top} \boldsymbol{\beta}_{j}.
\end{split}
\end{equation}
By combining equations \eqref{eq:Like}, \eqref{eq:BetaPrior}, and \eqref{eq:UtAU}, we derive the Gaussian conditional posterior distributions for $\boldsymbol{\beta}_{j}$ for $j = 2, \dots, J$ as:  
\begin{align*}
    &p\{\boldsymbol{\beta}_{j} \,|\, (\boldsymbol{\beta}_{\ell})_{\ell=1}^{j-1},  \cdot \} \\
    &\propto \exp\Big\{ -\frac{1}{2} \boldsymbol{\beta}_{j}^{\top} \boldsymbol{\Pi}_{j} \boldsymbol{\beta}_{j} \Big\}  \exp\Big\{- \frac{1}{2} \sum_{i=1}^{n}   T_{i} \lambda_{ij}^{-1} \Big(-2\boldsymbol{A}_{i,j,1:(j-1)} \mathbf{U}_{i,1:(j-1),1:(j-1)} \Tilde{\boldsymbol{x}}_{i}^{\top} \boldsymbol{\beta}_{j} \\
    &\qquad + \boldsymbol{\beta}_{j}^{\top}\Tilde{\boldsymbol{x}}_{i} \mathbf{U}_{i,1:(j-1),1:(j-1)}^{\top} \boldsymbol{A}_{i,1:(j-1),1:(j-1)} \mathbf{U}_{i,1:(j-1),1:(j-1)} \Tilde{\boldsymbol{x}}_{i}^{\top} \boldsymbol{\beta}_{j} \Big) \Big\} \\
    &\propto \mathcal{N}\Big(\boldsymbol{C}_{j}^{-1}\boldsymbol{\mu}_{j}, \boldsymbol{C}_{j}^{-1}\Big).
\end{align*}
This Gaussian conditional posterior distribution leads to the posterior sampling formulation for $\boldsymbol{\beta}_{j}$, as given in Equation \eqref{eq:SampBeta}.
\end{proof}

\subsection{Gibbs sampling: Full conditional distribution of \texorpdfstring{$\pi_{j\ell}$}{Lg}}

From the prior specification of the regression coefficients in equation \eqref{eq:AllBetaPriors}, we derive the conditional posterior distribution for the mixing probabilities $\boldsymbol{\pi} = (\boldsymbol{\pi}_{2}, \dots, \boldsymbol{\pi}_{J}) \in \mathbb{R}^{J(J-1)/2}$, where $\boldsymbol{\pi}_{j} := \boldsymbol{\pi}_{j,1:(j-1)} = (\pi_{j1}, \dots, \pi_{j,j-1}) \in \mathbb{R}^{j-1}$ denotes a vector of mixing probabilities. For $\ell < j$, $j = 2, \dots, J$, the conditional posterior probabilities under the cases $\pi_{j\ell} = 0$ and $\pi_{j\ell} = 1$ are given by:
\begin{align*}
    &p(\pi_{j\ell} = 0 \,|\, \cdot) \propto p(\pi_{j\ell} = 0)p(\boldsymbol{\beta}_{pj\ell} \,|\, \pi_{j\ell} = 0) \propto (1-q_{1}) \, \mathcal{N}(0, \tau_{0}^{2}), \\
    &p(\pi_{j\ell} = 1 \,|\, \cdot) \propto p(\pi_{j\ell} = 1)p(\boldsymbol{\beta}_{pj\ell} \,|\, \pi_{j\ell} = 1) \propto q_{1} \, \mathcal{N}(0, \tau_{1}^{2}) .
\end{align*}
Using these expressions, the conditional posterior probability for $\pi_{j\ell}$ is computed as: 
\begin{align*}
    &q_{1}^{\star} = \frac{q_{1} \mathcal{N}(\beta_{pj\ell}; 0, \tau_{1}^{2})}{(1-q_{1}) \mathcal{N}(\beta_{pj\ell}; 0, \tau_{0}^{2}) + q_{1} \mathcal{N}(\beta_{pj\ell}; 0, \tau_{1}^{2}) }, \\
    &p(\pi_{j\ell} \,|\, \cdot) \sim \text{Ber}(q_{1}^{\star}).
\end{align*}
Thus, posterior samples of $\pi_{j\ell}$ are drawn using the Gibbs sampler from a Bernoulli distribution with success probability $q_{1}^{\star}$, which reflects the relative contributions of the spike and slab components.

\section{Numerical illustrations using simulated data} \label{sec:sim}

We investigate the performance of our estimation approach through simulations designed to reflect the characteristics of ABIDE. We consider two settings with different sparsity levels and data dimensions. For both simulations, the sample consists of $n=500$ participants, each observed over $T_{i} = 200$ time points. The block partition vector $\boldsymbol{d}_{i}$ is generated using a multinomial distribution, ensuring that voxels are assigned approximately evenly across the ROIs. 

In the first simulation, we examine the effect of varying sparsity levels in a high-dimensional setting, where the number of voxels is $M_{i} = 5{,}000$ and the number of ROIs is set to $J = 50$. We consider three sparsity scenarios for the $p$-th covariate in the covariance matrix $\boldsymbol{\Sigma}_i$: highly sparse (95\% sparsity), moderately sparse (80\% sparsity), and relatively sparse (65\% sparsity). This means that the true values of $\boldsymbol{\pi} \in \mathbb{R}^{J(J-1)/2}$ are drawn from a Bernoulli distribution with a success probability chosen to yield a covariance matrix $\boldsymbol{\Sigma}_i$ exhibiting the specified sparsity level for the $p$-th covariate. For example, in the moderately sparse setting, 80\% of voxel pairs within the covariance matrix satisfy $\partial \boldsymbol{\Sigma}_i / \partial x_{ip} = 0$.   
In the second simulation, we investigate a very high-dimensional setting where the number of voxels is $M_{i} = 10{,}000$ and the number of ROIs is set to $J = 100$ or $J = 200$, resulting in data dimensions that are smaller than but comparable to those of ABIDE. We assume a moderate sparsity structure, ensuring that the covariance matrix $\boldsymbol{\Sigma}_i$ exhibits 80\% sparsity for the $p$-th covariate. 

The remaining parameters are specified as follows. The true values of participant-specific block-wise error terms $\eta_{ij}$ are independently sampled from a uniform sequence with spacing of $0.05$ between $0.05$ and $1.5$. The true values of the scaling factors of the Cholesky decomposition $\lambda_{ij}$ are set to $1/j$ for all $i=1,\dots,n$. The true regression coefficients $\beta_{qj\ell}$ are independently drawn from a standard Gaussian distribution for all $q = 1,\dots,p-1$ and $\ell < j$, where $j=2,\dots,J$. For the $p$-th covariate, we set $\beta_{pj\ell} = 2\mathbbm{1}(\pi_{j\ell} = 1)$. The participant-specific covariates $\boldsymbol{x}_{i}$ include $p=3$ variables, including an intercept term $x_{i1}$, a Bernoulli-distributed covariate $x_{i2}$ with a success probability of $0.5$, and a uniformly distributed covariate $x_{i3}$ ranging from $-0.5$ to $0.5$. 

We next describe the prior distributions. The priors for $\eta_{ij}$ and $\lambda_{ij}$ are both $\text{Inv-Gamma}(2.01, 1.01)$. The hyperparameter for the mixing probabilities is set to $q_{1}=0.5$, resulting in a prior distribution of $\pi_{j\ell} \sim \text{Bern}(0.5)$, which ensures that each $\beta_{pj\ell}$ has an equal prior probability of inclusion. The hyperparameters of the regression coefficients are specified as $\tau_{2}^{2} = 1$, $\tau_{1}^{2} = 1$, and $\tau_{0}^{2} = 0.01$. 

Results are evaluated based on $50$ simulated data sets, with the model in each replicate estimated with $6{,}000$ MCMC iterations, including a burn-in period of $1{,}000$ iterations. All simulations and data analyses are conducted on a SLURM server running Ubuntu 20.04.6 LTS, with an Intel Xeon Gold 6248 CPU (2.50 GHz, 1 core per task). The software environment included R 4.2.1 with Intel Math Kernel Library (MKL) for optimized linear algebra operations, as well as Rcpp and RcppArmadillo for high-performance computations and seamless integration of C++ code.  

We emphasize that both simulations involve an extremely high-dimensional setting in terms of both data size and the number of parameters to be sampled. The total number of between-voxel correlations is given by $\sum_{i=1}^{n} M_{i}^{2}$, which amounts to $1.25 \times 10^{10}$ for the first simulation and $5 \times 10^{10}$ for the second. Despite the massive scale of the data, our implementation of the BlocR model significantly reduces the number of parameters while maintaining flexibility. The parameters to be sampled include $\boldsymbol{\eta} \in \mathbb{R}^{nJ}$, $\boldsymbol{\lambda} \in \mathbb{R}^{nJ}$, $\boldsymbol{\beta} \in \mathbb{R}^{pJ(J-1)/2}$, and $\boldsymbol{\pi} \in \mathbb{R}^{J(J-1)/2}$. The parameter space, however, remains substantial. For $J = 50$, the number of parameters to be sampled is $54{,}900$; for $J = 100$, it increases to $119{,}800$; and for $J = 200$, it reaches $279{,}600$. As $J$ increases, computational complexity grows accordingly, leading to longer computation times, which are nonetheless reasonable given the large number of parameters being sampled.

Table \ref{table:sim1} presents the results from the first simulation, reporting the coverage rates of 95\% credible intervals for key quantities of interest, averaged over 50 simulations. Coverage is consistently maintained across all sparsity scenarios for the block-wise error terms $\boldsymbol{\eta}$ and the regression coefficients $\{ \beta_{1:(p-1),j\ell} \}_{\ell < j, j=2}^{J}$. Coverage for the mixing parameters $\boldsymbol{\pi}$ is assessed by verifying whether the true values match the median posterior probability after rounding, given that the mixing parameter is binary. Coverage is achieved in the highly sparse (95\%) and moderately sparse (80\%) settings, while undercoverage is observed in the relatively sparse (65\%) scenario. This outcome is expected, as prior studies have shown that continuous spike-and-slab posteriors for high-dimensional linear regression models tend to perform best in high-sparsity settings \citep{narisetty2014bayesian, chen2019fast}. An important quantity in this study is the coverage rate for the partial derivative $\partial \boldsymbol{\Sigma}_i / \partial x_{ip}$, which measures the sensitivity of the covariance structure to changes in the $p$-th covariate. Coverage is well maintained in the highly sparse and moderately sparse settings, while slight undercoverage is observed in the relatively sparse case. However, the coverage remains within an acceptable range. The computation time per iteration averages 6 seconds across all sparsity scenarios.

\begin{table}[ht!]
%\small
\centering
\begin{tabular}{ccccc}
& \multicolumn{4}{c}{coverage rate (se)} \\
\cline{2-5}
Sparsity & $\boldsymbol{\eta}$ & $\{\beta_{1:(p-1),j\ell}\}_{\ell<j,j=2}^{J}$ & $\boldsymbol{\pi}$ & $\partial \boldsymbol{\Sigma}_i/ \partial x_{ip}$ \\ \hline
95 \%  & 0.95 ($<$ 0.01) & 0.93 (0.04) & 0.96 (0.01) & 0.94 (0.05) \\ 
80 \%  & 0.95 ($<$ 0.01) & 0.95 (0.03) & 0.94 (0.01) & 0.94 (0.06) \\ 
65 \%  & 0.95 ($<$ 0.01) & 0.95 (0.04) & 0.73 (0.01) & 0.84 (0.06) \\ \hline
\end{tabular}
\caption{First simulation with $M=5{,}000$, $J=50$. Coverage rates (standard errors) of 95\% credible intervals for quantities of interest averaged across 50 simulations, for three sparsity scenarios: highly sparse (95\%), moderately sparse (80\%), and relatively sparse (65\%). 
} \label{table:sim1}
\end{table}

Table \ref{table:sim2} presents results from the second simulation, which considers larger data dimensions ($M=10{,}000$) with $J=100$ and $J=200$ under the moderately sparse (80\%) setting. As in the first simulation, coverage rates of 95\% credible intervals for key quantities of interest are reported, averaged over 50 simulations. Coverage is consistently maintained across all parameters. The computation time per iteration increases with model complexity, averaging 1 minute for $J=100$ and 15 minutes for $J=200$. 

\begin{table}[ht!]
%\small
\centering
\begin{tabular}{cccccc}
& & \multicolumn{4}{c}{coverage rate (se)} \\
\cline{3-6}
$M_{i}$ & $J$ & $\boldsymbol{\eta}$ & $\{\beta_{1:(p-1),j\ell}\}_{\ell<j,j=2}^{J}$ & $\boldsymbol{\pi}$ & $\partial \boldsymbol{\Sigma}_i/ \partial x_{ip}$ \\ \hline
$10,000$ & 100  & 0.95 ($<$ 0.01) & 0.94 (0.03) & 0.94 (0.01) & 0.95 (0.03) \\ 
$10,000$ & 200  & 0.95 ($<$ 0.01) & 0.95 (0.04) & 0.93 (0.02) & 0.94 (0.05) \\ \hline
\end{tabular}
\caption{Second simulation with $M=10,000$ with $J=100$ and $J=200$ under the moderately sparse (80\%) setting. Coverage rates (standard errors) of 95\% credible intervals for quantities of interest averaged across 50 simulations.
} \label{table:sim2}
\end{table}

\begin{figure}[ht!]
    \centering
    \includegraphics[width=0.9\linewidth]{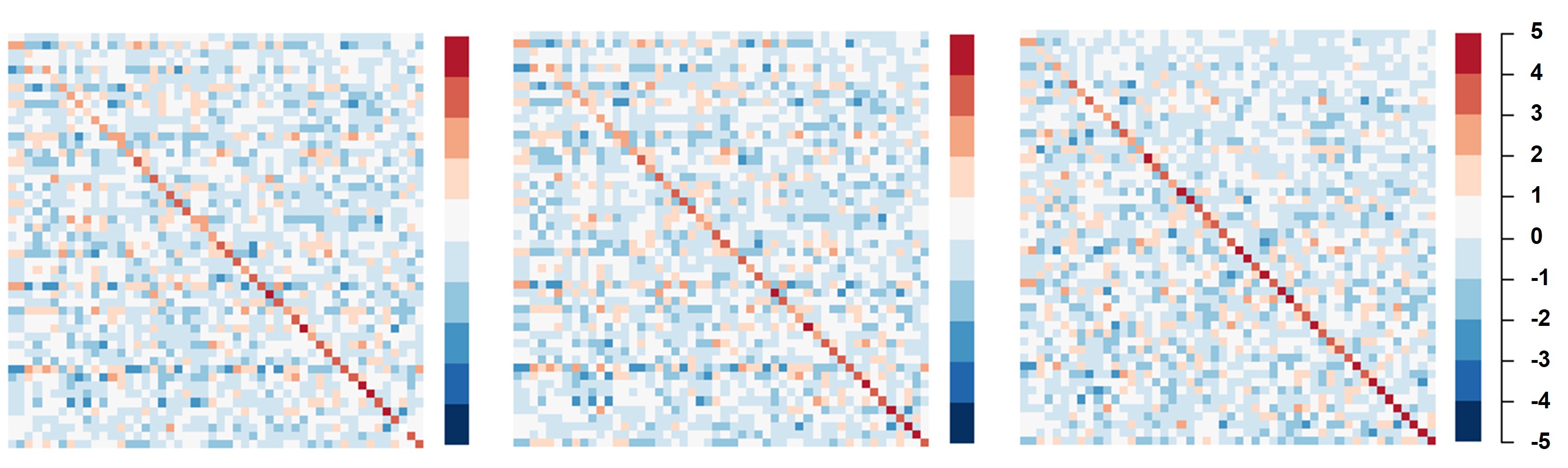}
    \caption{Comparison of the scaled $\boldsymbol{\Delta}_{i}$ from the true parameters (left), the scaled observed summary statistic $\boldsymbol{A}_{i}$ (middle), and the scaled estimated $\boldsymbol{\Delta}_{i}$ obtained from MAP parameter estimates (right) under the moderate sparsity scenario in the first simulation ($M_{i}=5,000$, $J=50$).}
    \label{fig:DeltaPlot}
\end{figure}

Figure \ref{fig:DeltaPlot} visually evaluates the performance of the BlocR model in estimating $\boldsymbol{\Delta}_{i}$, a key component of the covariance matrix. The plot compares the scaled $\boldsymbol{\Delta}_{i}$ computed from the true parameters (left), the scaled observed summary statistic $\boldsymbol{A}_{i}$ (middle), and the scaled estimated $\boldsymbol{\Delta}_{i}$ obtained from maximum a posteriori (MAP) parameter estimates (right) under the moderate sparsity scenario in the first simulation for a randomly selected participant. Due to the high-dimensional nature of the data, directly comparing entire covariance matrices is challenging. Instead, focusing on the key quantity $\boldsymbol{\Delta}_{i}$ allows for a more interpretable assessment. The results show that, by construction, the observed summary statistic $\boldsymbol{A}_{i}$ closely approximates the true $\boldsymbol{\Delta}_{i}$. Moreover, the estimated $\boldsymbol{\Delta}_{i}$ closely resembles the true values, demonstrating the model's ability to accurately recover the parameters. Note that all comparisons are based on the scaled versions of these quantities, as estimating the scaling factors $\boldsymbol{\lambda}$ is relatively more challenging. However, this does not impact the interpretability of the model.

\section{Identification of functional connectivity influenced by ASD status} \label{sec:dat}

\subsection{ABIDE preprocessing and covariate selection}

Our objective is to identify which voxel pairs' dynamic covariance is influenced by a participant's membership in either the Autism Spectrum Disorder (ASD) group or the typically developing (TD) control group.  Our parcellation is the hierarchical multi-resolution 17-network parcellation of \cite{Schaeferetal2018}, which contains $J=200$ ROIs. A majority voting approach was used to assign voxels to ROIs due to the difference in resolution between the atlas ($1\times 1 \times 1 \text{ mm}^{3}$) and the data ($3\times 3 \times 3 \text{ mm}^{3}$). After excluding voxels located outside the brain due to imperfections of the registration, each participant retains on average $M_{i} \approx 42{,}750$ voxels (with mean $42{,}750.45$ and standard deviation $1{,}591.21$ number of voxels). The number of voxels within each ROI, $d_{ij}$, has a mean of $213.75$ and a standard deviation of $95.24$.

\begin{figure}[ht!]
    \centering
    \includegraphics[width=0.9\linewidth]{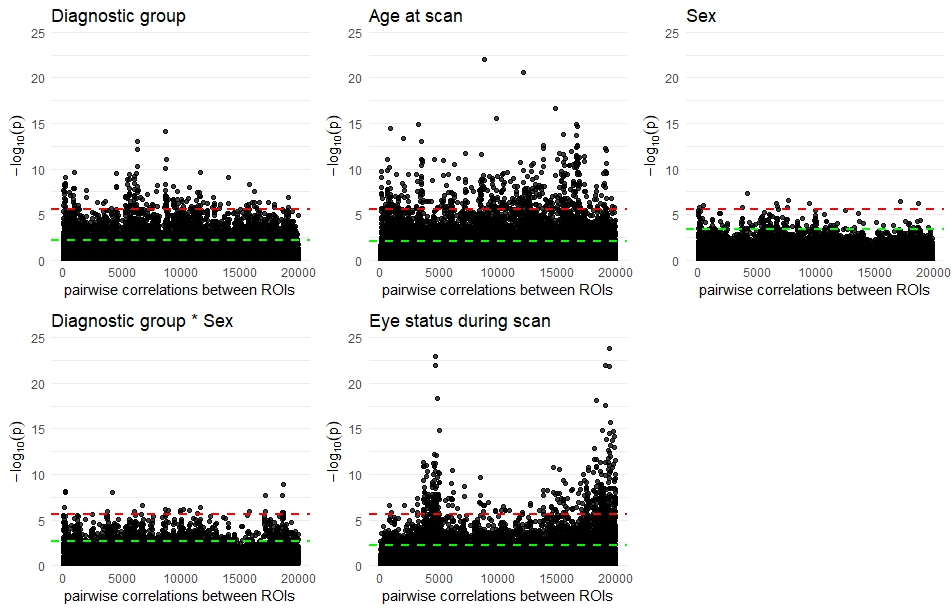}
    \caption{Manhattan plot showing the $p$-values from univariate regression of each of the selected covariates against the pairwise correlations between $J=200$ ROIs. The horizontal red line indicates the significance threshold using the Bonferroni correction. The horizontal green line represents the False Discovery Rate threshold calculated using the Benjamini-Hochberg procedure. Any points above these lines are considered statistically significant at the corresponding thresholds.}
    \label{fig:manhattan}
\end{figure}

We included only those participants who passed the functional quality assessment conducted through manual inspection by three independent raters. Additionally, participants were required to have at least one recorded voxel value for all ROIs. The covariates $\boldsymbol{x}_i \in \mathbb{R}^{p}$ include selected demographic and clinical characteristics of the participants. We chose candidate covariates based on previous studies \citep{qi2020common, heinsfeld2018identification}, focusing on phenotypic information relevant to ASD. The final covariate set includes $p=6$ variables: intercept, ASD status, age at the time of the scan, sex, group-sex interaction term, and eye status during the resting scan. These variables were selected based on an examination of Manhattan plots, which display the $p$-values from univariate regressions of each candidate covariate against pairwise correlations among ROIs. Figure \ref{fig:manhattan} displays the Manhattan plots for the selected variables, with Manhatten plots for other considered variables available in the Supplementary Material. As the primary objective is to identify voxels influenced by ASD or TD group membership, we assign the continuous spike-and-slab prior to the diagnostic group variable, treated as the $p$-th covariate. The covariates and outcomes $\boldsymbol{Y}_{i}$ are centered and scaled for analysis, making the covariance matrix effectively equivalent to the correlation matrix.

The total number of participants used in the analysis is $n=764$, with 374 in the ASD group and 390 in the TD group. A phenotypic summary of these participants based on the selected covariates is provided in Table \ref{tab:pheno}. For each participant, let $T_{i}$ denote the length of the rfMRI time series, which has a mean of $98.11$ and a standard deviation of $29.86$ after applying lag-2 thinning.  This thinning process helps ensure that the rfMRI outcomes are approximately independent. The total number of data points is $\sum_{i=1}^{n} M_{i}^{2} \approx 1.4 \times 10^{12}$. 

\begin{table}[ht!]
    \centering
    \begin{tabular}{lcccccccc}
        \hline
        & sample size & age & \multicolumn{2}{c}{sex} & IQ & \multicolumn{2}{c}{eye status} \\
        & & mean (sd) & male & female & mean (sd) & open & closed \\
        \hline
        ASD & 374 & 15.2 (6.4) & 330 & 44 & 104.3 (17.1) & 318 & 56 \\
        TD & 390 & 15 (5.8) & 307 & 83 & 110.6 (13) & 316 & 74 \\
        \hline
    \end{tabular}
    \caption{Phenotypic summary of ASD and TD groups for variables: age at the time of the scan, sex, full-scale IQ standard score, and eye status during the resting scan. Mean and standard deviation are provided for continuous variables, counts are reported for categorical variables.}
    \label{tab:pheno}
\end{table}

\subsection{Evaluating covariance matrix estimation}

\begin{figure}[ht!]
    \centering
    \includegraphics[width=1\linewidth]{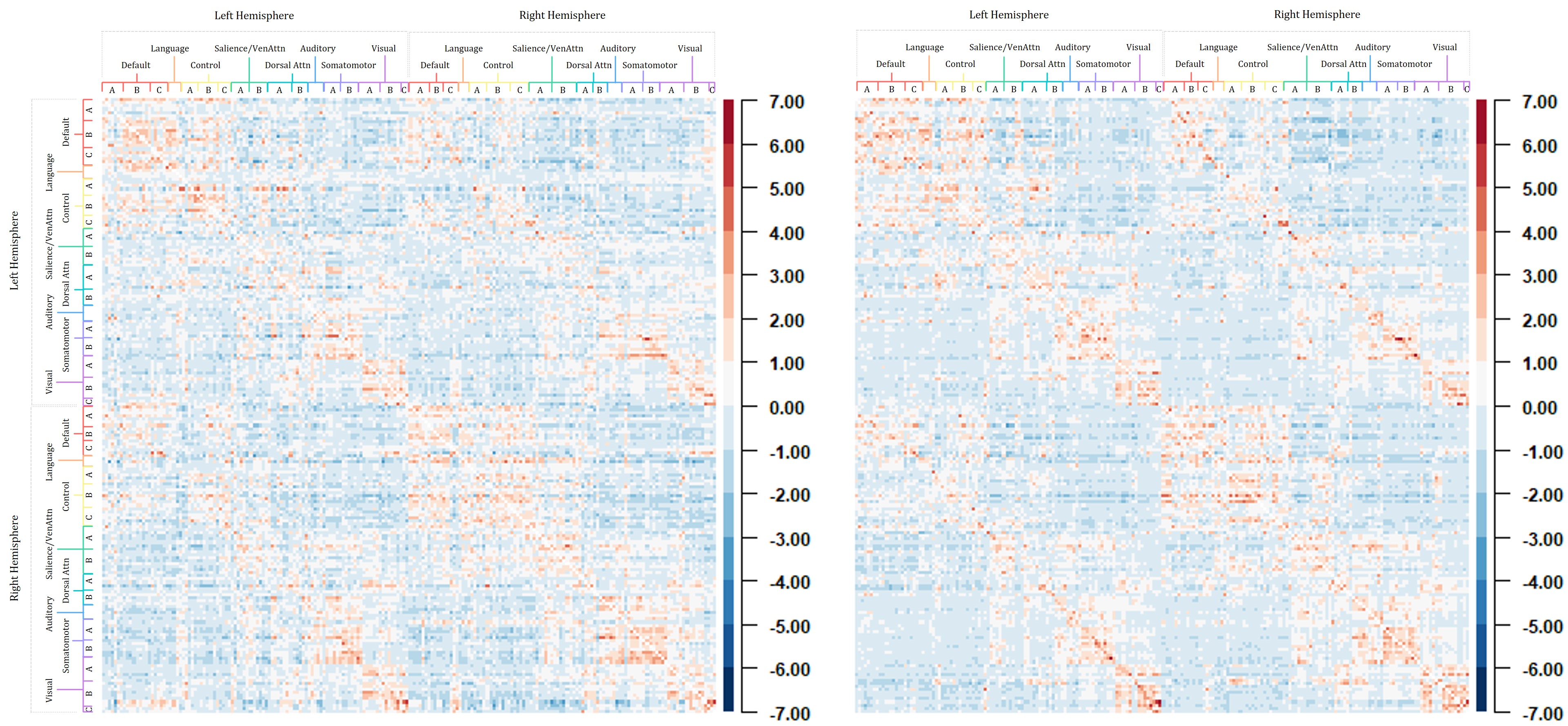}
    \caption{Comparison of the scaled observed summary statistic $\boldsymbol{A}_{i}$ (left) and the scaled estimated $\boldsymbol{\Delta}_{i}$ from MAP parameter estimates (right) for a randomly selected participant. The red banded structure along the diagonals and off-diagonals reflects the ROI ordering in the 17-network parcellation, where the first 100 ROIs correspond to the left hemisphere and the last 100 to the right hemisphere.}
    \label{fig:Delta_compare_plots}
\end{figure}

Results are based on $5{,}000$ MCMC iterations, including a burn-in period of $1{,}000$ iterations, with a computation time of $52$ minutes per iteration. To evaluate whether the model accurately estimates the covariance matrix, Figure \ref{fig:Delta_compare_plots} compares the scaled observed summary statistic $\boldsymbol{A}_{i}$ (left panel) with the scaled estimated $\boldsymbol{\Delta}_{i}$ obtained from MAP parameter estimates (right panel) for a randomly selected participant, as previously done in Section \ref{sec:sim}. A visual examination indicates that the two matrices are indeed similar.  

Both matrices exhibit patterns of positive values highlighted in red along the diagonals and off-diagonals, forming a banded structure. This pattern arises due to the ordering of ROIs in the 17-network parcellation: the first 100 ROIs correspond to the 17 networks in the left hemisphere, while the last 100 ROIs correspond to the 17 networks in the right hemisphere. A similar structure was observed in the ROI-level functional connectivity matrix shown in Figure \ref{fig:motiv}. It is important to clarify that while the ROI-level functional connectivity matrix represents the sample correlation matrix, the scaled $\boldsymbol{A}_{i}$ is a summary statistic of the sample covariance matrix, adjusted for the number of voxels within each ROI. Thus, the two matrices are related but not identical.

\subsection{Inference on functional connectivity and ASD status}

We conduct a hypothesis test to evaluate whether changes in ASD status $x_{ip}$ ($x_{ip} = 0$ for ASD group, $x_{ip} = 1$ for TD control) are associated with the covariance block $\boldsymbol{\Sigma}_{ij\ell}$ as specified by the model, for all $j, \ell = 1, \dots, J$. Since $x_{ip}$ is binary, we assess its effect on $\boldsymbol{\Sigma}_{ij\ell}$ by computing the discrete change when switching from $x_{ip} = 0$ to $x_{ip} = 1$, rather than using a partial derivative (which is more applicable for continuous covariates):
\begin{align}
    \label{eq:deriv_bin} &\boldsymbol{\Sigma}_{ij\ell}(x_{ip} = 1) - \boldsymbol{\Sigma}_{ij\ell}(x_{ip} = 0)\\ 
    &= \Big[ \frac{1}{\sqrt{d_{j}d_{\ell}}}  \sum_{\ell'=1}^{\ell} \lambda_{i\ell'} \Big\{ \beta_{pj\ell'}\beta_{p\ell\ell'} + \beta_{pj\ell'} \Big( \sum_{q=1}^{p-1} x_{iq}\beta_{q\ell\ell'} \Big) + \beta_{p\ell\ell'} \Big( \sum_{q=1}^{p-1} x_{iq}\beta_{qj\ell'} \Big)  \Big\}  \Big] \boldsymbol{1}_{d_{j} \times d_{\ell}}. \nonumber 
\end{align}
To assess statistical significance, we compute 95\% posterior credible intervals for equation \eqref{eq:deriv_bin}. If the credible interval does not contain zero, it provides strong evidence against the null hypothesis, indicating that voxels within the $(j,\ell)$-th ROI exhibit a significant association with changes in ASD status. 

\begin{figure}[ht!]
    \centering
    \includegraphics[width=0.5\linewidth]{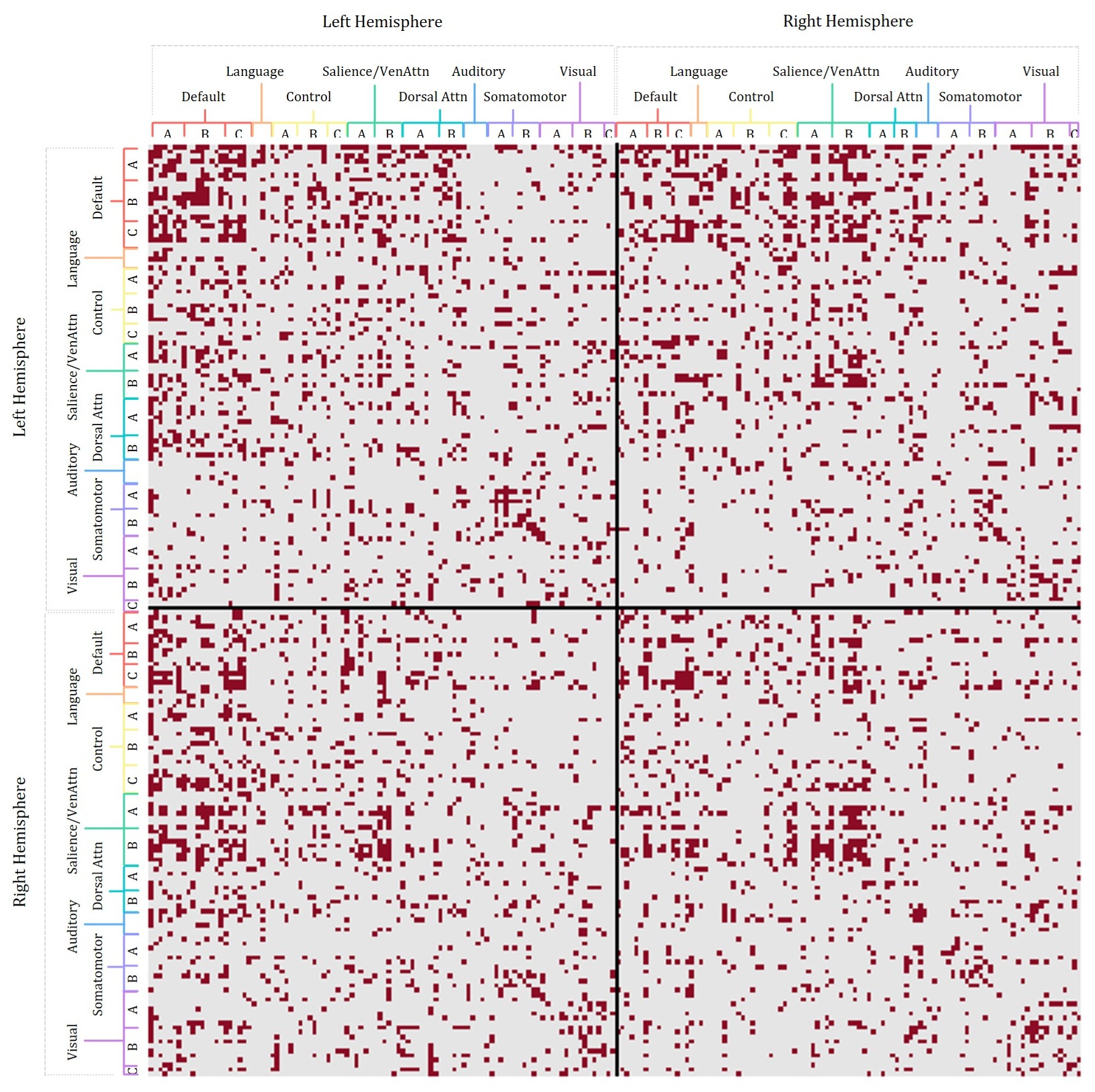}
    \caption{Covariance blocks (ROIs) shown in red indicate regions where changes in ASD status ($x_{ip}$) are statistically significant in over 95\% of participants. Black lines separate left and right hemispheres.}  \label{fig:covBlocks}
\end{figure}

We test our hypotheses across all 764 participants in the study. Our analysis identifies voxels within the $(j,\ell)$-th ROI that exhibit statistical significance in over 95\% of participants. These regions are particularly important as they reliably contribute to differentiating ASD from TD participants by capturing consistent differences in functional connectivity patterns. Figure \ref{fig:covBlocks} presents these results, revealing patterns of connectivity both within and between hemispheres. Notably, there appears to be a greater concentration of significant regions in the upper left and lower right quadrants across all hemispheric areas. 

\begin{figure}[ht!]
    \centering
    \includegraphics[width = 0.7\linewidth]{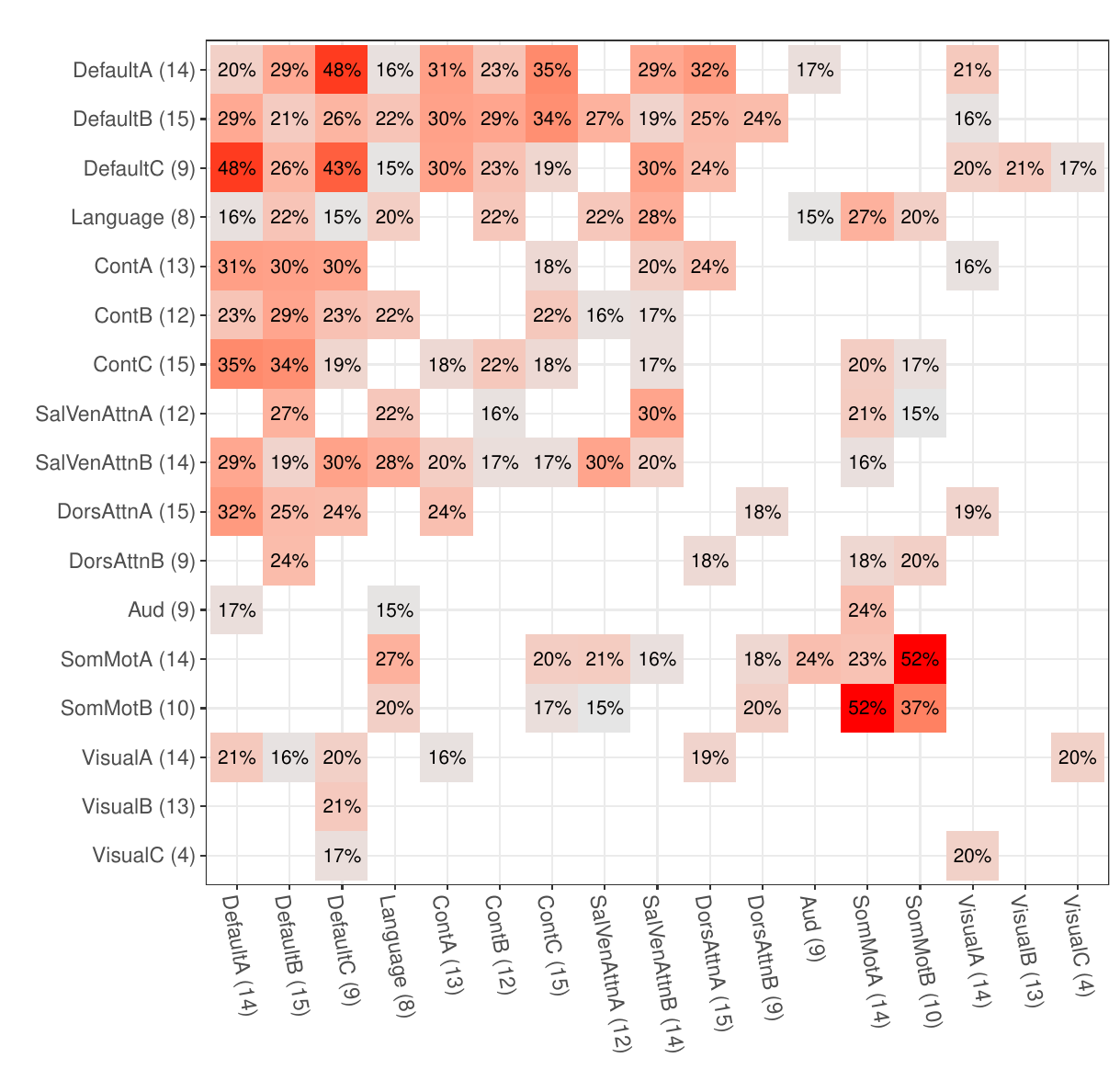}
    \caption{Percentage of ROIs in each network with statistically significant connections to the other 17 networks. Only ROIs with at least 15\% significant connections are shown. The number of ROIs per network is indicated in parentheses.} \label{fig:networks}
\end{figure}

Figure \ref{fig:networks} provides a network-level summary of our findings, illustrating the percentage of statistically significant ROIs in each network that connect to the other 17 networks. To enhance interpretability, we focus on ROIs where at least 15\% of connections are statistically significant. This visualization offers insights into broader functional connectivity differences associated with ASD.

Our findings align with previous research, supporting reported patterns of subnetwork and internetwork connectivity atypicalities in ASD. Specifically, the results highlight increased involvement of the default mode subnetworks \citep{padmanabhan2017default}, sensorimotor subnetworks \citep{coll2020sensorimotor}, and connectivity between the control and attention networks \citep{FarrantUddin2016}. These findings suggest that ASD-related patterns of atypical network connectivity are present in individuals with frequently co-occurring conditions. Future studies with larger, well-characterized ASD cohorts are needed to confirm these results.
%-----------------------------------------------------------------------------------------------%

\section{Conclusion} \label{sec:conc}

Although developed for large-scale neuroimaging data, the BlocR model has broader applications. Exploring its use in other domains, such as financial data, could be valuable. While block structures have been studied in finance \citep{archakov2022canonical}, they have not been examined at this scale nor have they incorporated covariates in the model. Future work should assess whether our inferential framework yields similarly robust insights in this context.

\bibliographystyle{plainnat}
\bibliography{arxiv-main}

\end{document}